# ANTICORRELATED CN AND CH VARIATIONS ON THE 47 TUCANAE MAIN-SEQUENCE TURNOFF

MICHAEL M. BRILEY[1]
McDonald Observatory and Department of Astronomy, University of Texas, Austin, Texas 78712

JAMES E. HESSER[1]
Dominion Astrophysical Observatory, Herzberg Institute of Astrophysics
National Research Council of Canada
5071 West Saanich Road, Victoria, British Columbia, V8X 4M6, Canada

R. A. BELL
Department of Astronomy, University of Maryland, College Park, Maryland 20742

MICHAEL BOLTE[1] AND GRAEME H. SMITH
University of California Observatories / Lick Observatory
Board of Studies in Astronomy and Astrophysics
University of California, Santa Cruz, California 95064

ABSTRACT

Observations of CN and CH band strengths among a random sample of main-sequence turn-off stars ($+3.9 < M_V < +4.6$) in the globular cluster 47 Tuc (NGC 104, C0021-723) were made with the CTIO Argus fiber spectrograph for the purpose of determining the ratio of CN-strong to CN-weak stars and investigating the behavior of CH relative to CN. Of the 20 turn-off stars, 12 were found to be CN-strong while 8 appear to be CN-weak. This ratio of CN-strong to CN-weak stars is similar to the ratios found among the more luminous 47 Tuc stars and implies little change in the overall distribution of CN with evolutionary state, although the present sample size is small. A general anticorrelation between CN and CH is also observed in that the CN-weak main-sequence stars all (with one possible exception) exhibit strong CH bands - a trend similar to that found among the brighter stars. That these variations occur among such relatively un-evolved stars and that the overall CN distribution appears to be independent of evolutionary state presents serious challenges to internal or mixing theories of their origin. We therefore suggest that at least some component of the C and N abundance inhomogeneities observed among the brighter (more evolved) stars of this cluster appears to have been established prior to the commencement of evolution up the red giant branch.
*Sbject headings:* clusters: globular - stars: abundances - stars: late-type

## 1. INTRODUCTION

Over the past two decades, considerable effort has been directed towards understanding the origins of the star-to-star abundance variations in light elements found in every globular cluster studied to date. Reviews of these works may be found in Smith (1987), Norris (1988), Smith (1989), Suntzeff (1989, 1993). The observations are generally interpreted within one of two contexts: 1) the variations are due to processes internal to the stars themselves (such as the deep dredge-up of material processed through a region of active nucleosynthesis, e.g., Sweigart & Mengel 1979, Langer & Kraft 1984, Smith & Tout 1992), or 2) the inhomogeneities are due to some "primordial" process (i.e., a process that was only active early in a cluster's history, such as might have been instigated by a generation of massive or intermediate mass stars, Cottrell & Da Costa 1981), that enriched the gas from which some of the low-mass cluster stars formed.

The Na and Al abundance variations observed among cluster red giants (c.f., Drake, Plez & Smith 1993), once thought to be strong evidence of primordial inhomogeneities (c.f., Peterson 1980; Cottrell & Da Costa 1981; Smith & Wirth 1991; Brown & Wallerstein 1992; Drake, Smith & Suntzeff 1992), may instead be explicable in terms of proton-addition nucleosynthesis within the hydrogen-burning shell of these giants (Langer, Hoffman & Sneden 1993). It therefore becomes important to search for CN variations among less evolved stars within which such nucleosynthesis is not thought to have been operative. This is necessary in order to determine whether the complicated patterns of CN, CH and [O/Fe] variations observed among cluster red giants can be entirely attributed to processes that have acted during the giant-branch phase of evolution or whether there is still some need to invoke "primordial" contributions to these inhomogeneities. Any inhomogeneities observed among main-sequence turnoff stars or unevolved main-sequence stars could not be attributed to processes that are only operative during the giant-branch phase of evolution.

---

[1]Visiting Astronomer, Cerro Tololo Inter-American Observatory, National Optical Astronomy Observatories, operated by AURA, Inc., under contract with the National Science Foundation.

The cluster upon which most effort to date has been exerted to search for CN variations among relatively unevolved stars is 47 Tuc, whose G4 spectral type (Hesser & Shawl 1985) is indicative of turnoff stars sufficiently cool and metal-rich that CN and CH molecular features, while weak, should be detectable. Following the photometry of main-sequence stars in 47 Tuc by Hesser & Hartwick (1977) and Hesser et al. (1987), the spectroscopy of stars in the turnoff region of the color-magnitude diagram (CMD) was reported by Hesser (1978), Hesser & Bell (1980), Bell, Hesser & Cannon (1983), and Briley, Hesser & Bell (1991). These investigations have demonstrated that CN differences do exist among both upper main-sequence and main-sequence turnoff stars in 47 Tuc. In addition to 47 Tuc, star-to-star CN variations have also been found among main-sequence stars in NGC 6752 by Suntzeff (1989). Despite the considerable amount of telescope time that has been invested in these programs, all of these studies are based on relatively small or biased samples, so the question has remained as to whether two of the main characteristics of CN variations that are observed on the red-giant branch, i. e., a bimodal CN distribution and a general CN/CH anticorrelation, are also evinced by the main-sequence stars. The advent of multi-object fiber spectrographs on 4-m class telescopes has made it worthwhile to return to such investigations. In this paper we report upon spectra of main-sequence turnoff stars in 47 Tuc obtained with the Argus fiber-fed spectrograph on the CTIO 4-m telescope in 1992

## 2. OBSERVATIONS

The spectroscopic observations were made under poor seeing conditions (> 2.5 arc seconds) on the nights of 18-20 September 1992 with the CTIO 4-m telescope and Argus fiber spectrograph. A sample of main-sequence turnoff stars was randomly chosen for spectroscopy from a montage of CCD images obtained in 1989 with the CTIO 4-m and 0.9-m telescopes. Program star coordinates, V magnitudes, and B–V colors were measured from these CCD images, and this information is listed in Table 1. The photometric data consist of co-added short exposure frames ($6 \times 40$ s in V and $6 \times 60$ s in B) for two overlapping fields taken with the CTIO 4-m PFCCD camera using TI#1 and a mosaic of several overlapping fields imaged with the CTIO 0.9-m CFCCD camera using TI#3 with typical exposure times of $5 \times 120$ s in V and $5 \times 200$ s in B. Calibration of the 4-m data was via transfer of measurements to Landolt (1983) and Graham (1982) photometric standard stars. The solution for the transformation coefficients was based on 80 or more measurements of standard stars on each of the three nights. The 0.9-m data were calibrated by determining color coefficients based on measures of 39 standard stars imaged through clouds during the run with the photometric zeropoints determined by comparison with stars common to the 4-m fields. The RMS scatter around the standard star solution was 0.008 mag in V and 0.011 mag in B.

TABLE 1:
Program Stars, Their Positions, and Colors

| Star | RA (1950) | | | Dec (1950) | | | Finder Location[1] | | V | (B-V) | σV | σ(B-V) | n |
|---|---|---|---|---|---|---|---|---|---|---|---|---|---|
| | | | | | | | X | Y | | | | | |
| 1 | 0 | 18 | 19.74 | −72 | 21 | 34.89 | −282.30 | 10.78 | 16.18 | 0.83 | 0.01 | 0.01 | 7.5 |
| 2 | 0 | 19 | 37.71 | −72 | 23 | 47.77 | 61.283 | 143.66 | 16.48 | 0.85 | 0.02 | 0.03 | 1 |
| 3 | 0 | 19 | 52.40 | −72 | 24 | 0.87 | 126.86 | 156.76 | 16.80 | 0.75 | 0.02 | 0.03 | 1 |
| 4 | 0 | 18 | 44.42 | −72 | 21 | 48.41 | −171.17 | 24.30 | 17.16 | 0.69 | 0.01 | 0.01 | 6 |
| 5 | 0 | 18 | 49.51 | −72 | 20 | 25.92 | −141.57 | −58.19 | 17.35 | 0.59 | 0.01 | 0.01 | 4 |
| 6 | 0 | 19 | 50.69 | −72 | 20 | 50.68 | 134.77 | −33.43 | 17.35 | 0.59 | 0.01 | 0.02 | 3 |
| 7 | 0 | 19 | 24.51 | −72 | 19 | 30.99 | 22.14 | −113.12 | 17.38 | 0.57 | 0.02 | 0.03 | 2 |
| 8 | 0 | 18 | 1.71 | −72 | 18 | 47.30 | −351.70 | −156.81 | 17.39 | 0.59 | 0.02 | 0.02 | 1 |
| 9 | 0 | 19 | 13.22 | −72 | 19 | 58.29 | −31.49 | −85.82 | 17.47 | 0.57 | 0.01 | 0.02 | 3.5 |
| 10 | 0 | 19 | 56.30 | −72 | 22 | 56.66 | 149.86 | 92.55 | 17.48 | 0.63 | 0.01 | 0.05 | 1 |
| 11 | 0 | 19 | 54.11 | −72 | 18 | 53.33 | 160.03 | −150.78 | 17.55 | 0.58 | 0.03 | 0.04 | 1 |
| 12 | 0 | 19 | 5.62 | −72 | 20 | 55.91 | −70.66 | −28.20 | 17.56 | 0.56 | 0.01 | 0.02 | 3.5 |
| 13 | 0 | 20 | 2.12 | −72 | 23 | 28.74 | 173.61 | 124.63 | 17.60 | 0.51 | 0.02 | 0.03 | 1 |
| 14 | 0 | 19 | 58.18 | −72 | 21 | 29.93 | 165.61 | 5.82 | 17.60 | 0.60 | 0.02 | 0.03 | 1 |
| 15 | 0 | 19 | 34.12 | −72 | 20 | 39.92 | 60.30 | −44.19 | 17.64 | 0.52 | 0.01 | 0.02 | 4 |
| 16 | 0 | 19 | 41.82 | −72 | 21 | 48.80 | 89.66 | 24.69 | 17.67 | 0.49 | 0.01 | 0.02 | 4 |
| 17 | 0 | 18 | 36.21 | −72 | 21 | 28.78 | −206.95 | 4.67 | 17.68 | 0.58 | 0.01 | 0.01 | 5.5 |
| 18 | 0 | 18 | 3.96 | −72 | 20 | 0.85 | −346.97 | −83.26 | 17.69 | 0.60 | 0.01 | 0.01 | 7 |
| 19 | 0 | 18 | 53.98 | −72 | 22 | 20.12 | −130.18 | 56.01 | 17.70 | 0.56 | 0.01 | 0.01 | 5 |
| 20 | 0 | 19 | 45.02 | −72 | 19 | 17.44 | 116.66 | −126.67 | 17.75 | 0.63 | 0.02 | 0.04 | 1 |
| 21 | 0 | 19 | 1.66 | −72 | 21 | 31.47 | −91.46 | 7.36 | 17.76 | 0.55 | 0.01 | 0.01 | 3.5 |
| 22 | 0 | 18 | 37.37 | −72 | 20 | 36.42 | −197.60 | −47.69 | 17.80 | 0.53 | 0.01 | 0.01 | 5.5 |
| 23 | 0 | 19 | 20.97 | −72 | 21 | 41.62 | −4.50 | 17.51 | 17.80 | 0.56 | 0.01 | 0.01 | 5.5 |
| 24 | 0 | 18 | 23.52 | −72 | 20 | 31.59 | −260.26 | −52.52 | 17.92 | 0.56 | 0.01 | 0.01 | 8.5 |

[1] In arcsec, see Figure 1.

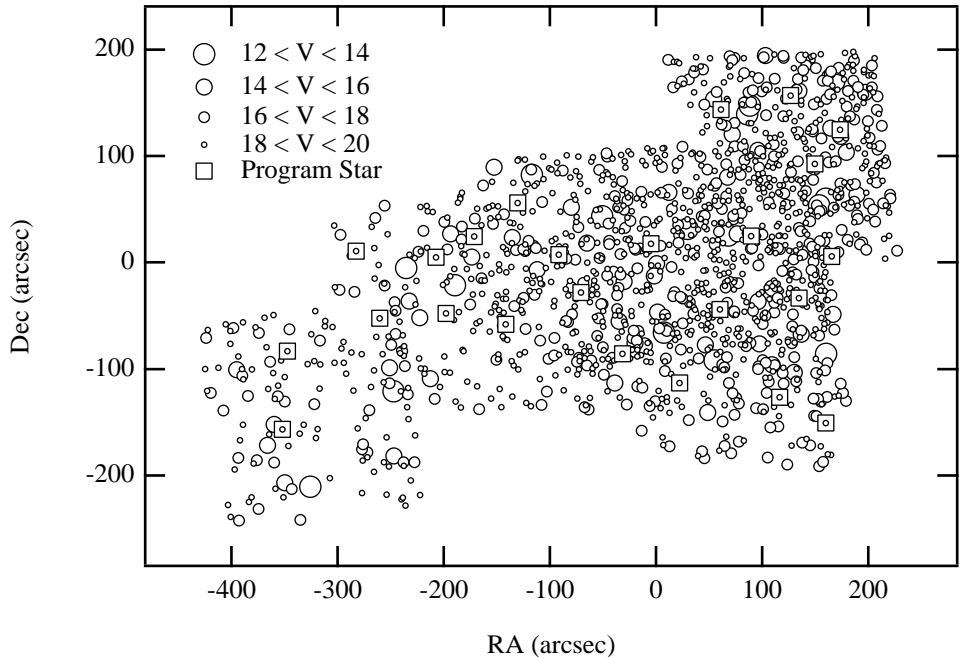

Figure 1: A finding chart for the program stars produced from a mosaic of CCD frames. The field center is roughly 11 arc min from the cluster center.

Table 1 lists right ascensions, declinations, (x,y) positions, the standard deviations about the means of the individual measurements, as well as the number of observations of each spectroscopic program star. A finding chart for these stars is given in Figure 1, overlaid on a schematic of the fields for which the photometry was derived. The locations of these stars on the fiducial CMD derived from the same data set are plotted in Figure 2. There are four stars included in the Argus field which are located near the base of the red-giant branch, while the remaining stars are located near the main-sequence turnoff with magnitudes in the range $17.3 < V < 18.0$ ($+3.9 < M_V < +4.6$; the apparent distance modulus was taken to be $(m-M)_V = 13.40$ and $E(B-V) = 0.04$, Hesser et al. 1987). Our program stars straddle the bluest part of the main-sequence turnoff region, which is located near V ~ 17.7.

Argus has 24 movable arms at the 4-m telescope's prime focus that can cover a field 50 arc minutes in diameter. Attached to each of these arms are two 100 μm (1.8 arc sec) diameter fibers, one at the arm tip and the other 1.7 mm (30 arc sec) outward, which are used to obtain object and sky spectra, respectively. Light from the fibers is fed to a bench-mounted spectrograph located at the base of the telescope. A more detailed description of the system can be found in the CTIO Users Manual and NOAO Newsletters 16 and 20. The program stars were each assigned to one of the twenty four fibers constituting a single Argus field.

The spectrograph was used in the RC-mode with the Reticon CCD detector in its unbinned format of 1200 x 400 pixels, each pixel being 27 μm on a side. The KPGL-1 grating (632 lines mm$^{-1}$, blazed at 4200 Å) provided spectral coverage from 3500 to 5700 Å at a dispersion of 1.82 Å pix$^{-1}$. The resulting resolution was 2.3 pix (4.2 Å) FWHM as determined from arc spectra. In order to minimize read noise, the gain was set to maximum, providing 0.3 electrons per analog-digital unit and a read noise of 3.1 electrons. Exposure times were generally 3600 s, with multiple exposures being taken on each night, although seeing variations resulted in considerable exposure-to-exposure differences in stellar flux.

TABLE 2
Definitions of Index Bandpasses

| Bandpass | Start Wavelength (Å) | End Wavelength (Å) |
|---|---|---|
| $F_{\lambda,F38}$ | 3846 | 3883 |
| $F_{\lambda,C39}$ | 3883 | 3916 |
| $F_{\lambda,F43}$ | 4280 | 4320 |
| $F_{\lambda,C42}$ | 4220 | 4280 |

Standard bias and dark exposures were made at the start and end of each night. A He-Ar lamp was observed for the wavelength calibrations; one long exposure through each fiber was made at the start and end of each night to determine the fiber-to-fiber differences in dispersion. For the program observations, an arc was observed at the beginning of every field change, to check for shifts in the dispersion. The bench-mounted spectrograph is quite stable and this proved more than adequate for the wavelength calibrations. Flats were also taken of a white spot on the dome interior that was illuminated by an incandescent lamp, with each fiber being extended inward, toward the field center, as far as possible. These exposures, plus twilight flats of the solar spectrum, were then used to normalize the detector/fiber responses. Extraction of the spectra followed standard procedures using the NOAO IRAF image reduction package (V2.10) and the Doargus script set. Because the seeing exceeded the size of the fibers in the image plane, flux standards were not observed and the final program spectra retain the instrumental response function (they have not been flux calibrated).

## 3. ANALYSIS AND RESULTS

### 3.1. INDICES

The strengths of absorption due to the 3883 Å CN and 4300 Å CH (G) bands were measured via a set of spectroscopic indices, S(3839) and $s_{CH}$, respectively, defined as:

$$S(3839) = -2.5 \log (F_{\lambda,F38} / F_{\lambda,C39}),$$

and,

$$s_{CH} = -2.5 \log ( F_{\lambda,F43} / F_{\lambda,C42}),$$

where $F_\lambda$ refers to the mean counts over the wavelength intervals specified in Table 2. These indices are basically measures of the flux blocked by a molecular feature in units of magnitude. The average indices measured from multiple exposures are listed in Table 3, along with the derived standard deviations ($\sigma$). Note that some exposures were not included in the calculation of the indices as cosmic ray events occurred within the index bandpasses. The number of exposures used in the statistics are also given in Table 3 (in the columns headed nS(3839) and ns$_{CH}$).

TABLE 3:
Indices Measured from Program Stars

| Star | V | B–V | S(3839) | σ[S(3839)] | δS(3839) | n[S(3839)] | $s_{CH}$ | σ[$s_{CH}$] | δ$s_{CH}$ | n[$s_{CH}$] |
|---|---|---|---|---|---|---|---|---|---|---|
| 1 | 16.18 | 0.83 | 0.223 | 0.041 | 0.028 | 5 | 0.088 | 0.022 | –0.075 | 5 |
| 2 | 16.48 | 0.85 | 0.512 | 0.029 | 0.297 | 5 | 0.061 | 0.010 | –0.101 | 5 |
| 3 | 16.80 | 0.75 | 0.083 | 0.083 | –0.095 | 2 | 0.057 | 0.007 | –0.104 | 2 |
| 4 | 17.16 | 0.69 | 0.477 | 0.052 | 0.474 | 6 | 0.083 | 0.024 | –0.069 | 5 |
| 5 | 17.35 | 0.59 | 0.061 | 0.033 | 0.154 | 5 | 0.064 | 0.011 | –0.054 | 5 |
| 6 | 17.35 | 0.59 | 0.308 | 0.045 | 0.400 | 5 | 0.015 | 0.010 | –0.103 | 5 |
| 7 | 17.38 | 0.57 | 0.241 | 0.050 | 0.336 | 5 | –0.014 | 0.012 | –0.130 | 5 |
| 8 | 17.39 | 0.59 | 0.188 | 0.031 | 0.283 | 4 | 0.036 | 0.017 | –0.080 | 5 |
| 9 | 17.47 | 0.57 | 0.015 | 0.050 | 0.116 | 5 | 0.046 | 0.027 | –0.065 | 5 |
| 10 | 17.48 | 0.63 | 0.177 | 0.070 | 0.278 | 5 | –0.016 | 0.034 | –0.127 | 7 |
| 11 | 17.55 | 0.58 | 0.144 | 0.037 | 0.248 | 4 | 0.026 | 0.029 | –0.082 | 4 |
| 12 | 17.56 | 0.56 | –0.013 | 0.052 | 0.091 | 5 | 0.036 | 0.031 | –0.071 | 5 |
| 13 | 17.60 | 0.51 | 0.091 | 0.046 | 0.196 | 4 | –0.027 | 0.015 | –0.133 | 5 |
| 14 | 17.60 | 0.60 | –0.029 | 0.083 | 0.076 | 4 | 0.032 | 0.020 | –0.074 | 4 |
| 15 | 17.64 | 0.52 | 0.183 | 0.071 | 0.289 | 4 | –0.025 | 0.013 | –0.130 | 5 |
| 16 | 17.67 | 0.49 | –0.005 | 0.067 | 0.102 | 4 | –0.014 | 0.054 | –0.118 | 7 |
| 17 | 17.68 | 0.58 | 0.140 | 0.050 | 0.247 | 5 | 0.027 | 0.034 | –0.076 | 5 |
| 18 | 17.69 | 0.60 | 0.259 | 0.084 | 0.366 | 5 | 0.025 | 0.013 | –0.078 | 4 |
| 19 | 17.70 | 0.56 | 0.164 | 0.019 | 0.272 | 6 | 0.025 | 0.029 | –0.078 | 6 |
| 20 | 17.75 | 0.63 | 0.220 | 0.010 | 0.327 | 2 | 0.044 | 0.019 | –0.059 | 2 |
| 21 | 17.76 | 0.55 | 0.262 | 0.029 | 0.369 | 4 | 0.028 | 0.022 | –0.075 | 6 |
| 22 | 17.80 | 0.53 | 0.068 | 0.076 | 0.174 | 4 | 0.064 | 0.018 | –0.039 | 4 |
| 23 | 17.80 | 0.56 | –0.008 | 0.044 | 0.098 | 4 | 0.094 | 0.025 | –0.009 | 4 |
| 24 | 17.92 | 0.56 | 0.061 | 0.099 | 0.165 | 5 | 0.051 | 0.043 | –0.055 | 4 |

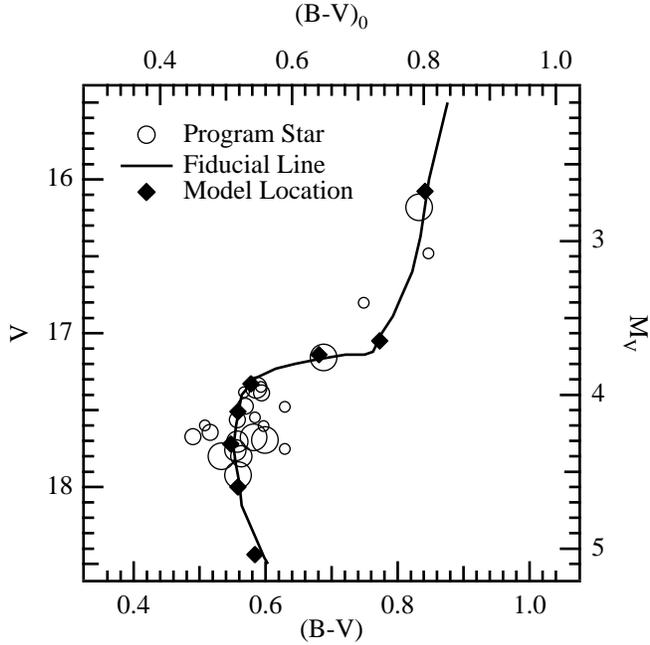

Figure 2: The location of the 47 Tuc program stars in the color-magnitude diagram measured from the photometric data reported in this paper. The size of the circles is proportional to the errors in the photometry (although not drawn to scale - see Table 1). Also plotted are the locations of the model atmosphere points and the fiducial locus for 47 Tuc derived from the present data set.

## 3.2 RESULTS

The main results of this investigation are summarized in Figures 3, and 6 - 8, which illustrate the behavior of the CN and CH bands among the main-sequence turnoff (17.3 < V < 18.0) program stars. The CN band strength index, S(3839), is plotted for these stars as a function of V and absolute magnitude in Figure 3. Two sequences of stars can be delineated: a CN-weak group with S(3839) < 0.09 (open circles) and a group with S(3839) > 0.09 having enhanced 3883 Å CN bands (filled circles). The main-sequence turnoff reaches it's bluest color at V ≈ 17.7 (M$_V$ ≈ 4.3), and as is illustrated by Figure 3, the CN index for both CN-enhanced and CN-weak stars appears to pass through a minimum near this magnitude. A typical difference in CN index between the CN-weak and CN-enhanced groups of stars in Figure 6 is $\Delta$S(3839) ≈ 0.2. A similar dispersion in CN band strengths was observed among a (biased) sample of stars in the same magnitude range by Briley et al. (1991) that were selected in expectation that they would sample the extrema of CN band strengths.

The existence of the CN-weak and CN-enhanced groups of main-sequence stars can be further demonstrated by plotting a histogram of S(3839) residuals which are measured with respect to a baseline in the S(3839) versus V diagram that would be defined by stars of C and N in the solar ratio. The baseline used in this analysis was obtained by interpolation between a set of theoretical S(3839) indices computed from synthetic spectra appropriate to several locations along the 47 Tuc giant-branch and turnoff regions.

A basic grid of stellar atmosphere parameters for these regions is given in Briley et al. (1991), to which we have added four additional points both to improve the model coverage and to correct for a 0.13 mag error in their applied distance modulus (note that this correction has little effect on their results). The procedures and assumptions involved in selecting the proper effective temperatures and surface gravities are outlined in Briley et al. (1991). The final grid of model parameters is given in Table 4.

TABLE 4:
Model Atmosphere Parameters and Resulting Indices

| T$_{eff}$ | log g | M$_v$ | V[3] | (B–V)$_0$ | B–V[3] | Solar CNO ratio | | CN-cycle[1] | | CN + ON-cycle[2] | |
|---|---|---|---|---|---|---|---|---|---|---|---|
| | | | | | | S(3839) | s$_{CH}$ | S(3839) | s$_{CH}$ | S(3839) | s$_{CH}$ |
| 5000 | 3.25 | 2.677 | 16.077 | 0.802 | 0.842 | 0.263 | 0.163 | 0.370 | 0.145 | 0.571 | 0.143 |
| 5200 | 3.7 | 3.646 | 17.046 | 0.733 | 0.773 | 0.149 | 0.160 | 0.249 | 0.139 | 0.427 | 0.134 |
| 5475 | 3.85 | 3.74 | 17.140 | 0.641 | 0.681 | 0.013 | 0.155 | 0.091 | 0.128 | 0.223 | 0.118 |
| 5820 | 4.05 | 3.927 | 17.327 | 0.538 | 0.578 | –0.091 | 0.119 | –0.055 | 0.090 | 0.008 | 0.079 |
| 5900 | 4.15 | 4.111 | 17.511 | 0.518 | 0.558 | –0.103 | 0.109 | –0.073 | 0.080 | –0.023 | 0.070 |
| 5950 | 4.25 | 4.322 | 17.722 | 0.507 | 0.547 | –0.108 | 0.102 | –0.082 | 0.075 | –0.038 | 0.065 |
| 5925 | 4.35 | 4.595 | 17.995 | 0.518 | 0.558 | –0.102 | 0.107 | –0.072 | 0.079 | –0.023 | 0.068 |
| 5850 | 4.5 | 5.036 | 18.436 | 0.544 | 0.584 | –0.085 | 0.118 | –0.047 | 0.088 | 0.018 | 0.077 |

[1] [C/A] = –0.21, [N/A] = +0.41: The maximum CN absorption possible via the CN–cycle, starting from a solar CN ratio (see Smith & Bell 1986).
[2] [C/A] = –0.3, [N/A] = +0.79, [O/A] = –0.2: The result of moderate mixing into a region of both CN and ON–processing.
[3] Assuming (m–M)$_V$ = 13.40 and E(B–V) = 0.04 (see Hesser et al. 1987).

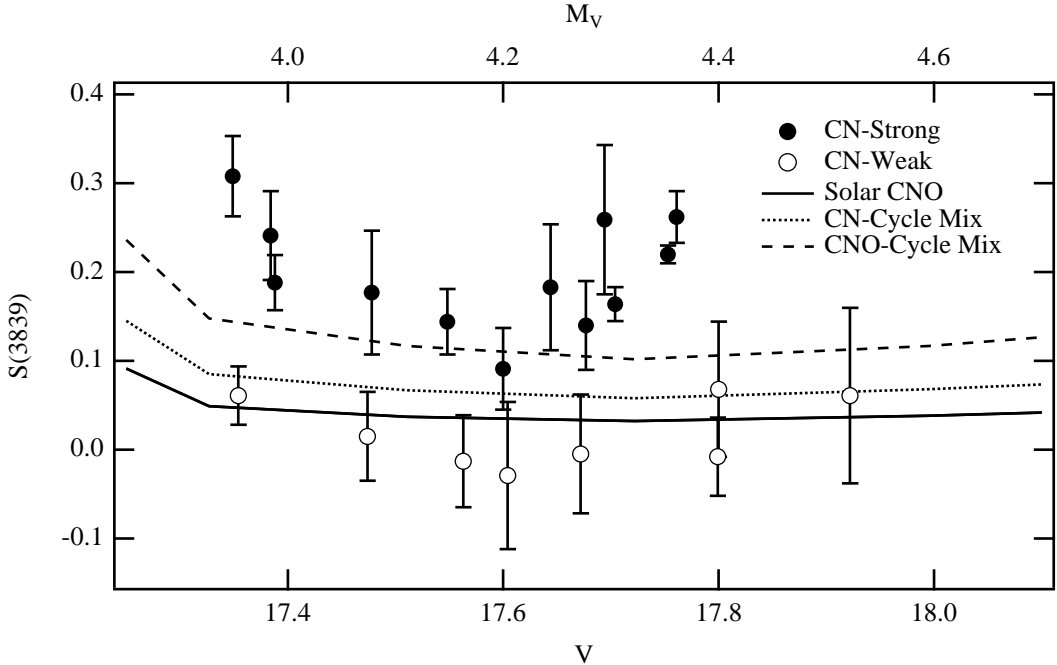

Figure 3: CN band strength indices, S(3839), are plotted for the main-sequence program stars (V > 17.3). Of the 20 stars, 12 appear to have enhanced CN-band strengths. Also plotted are indices computed from model atmospheres with differing CNO abundances, to which a zero-point shift of +0.14 has been applied (see text).

We have re-computed the synthetic spectra with the SSG program (Bell & Gustafsson 1978) using atomic and molecular data which have been improved over those of our previous investigations. Details of the changes are given in Bell, Paltoglou & Tripicco (1994, BPT). The line list adopted in this work is referred to by BPT as the "N" list. The most important difference, as far as this paper is concerned, is the adoption of a slightly larger dissociation energy for CN (7.65 eV as given in Bauschlicher, Langhoff & Taylor 1988, rather than the 7.50 eV used previously). New transition probabilities for the CN violet system have also been taken from Bauschlicher et al. (1988). BPT tested the improvements in their synthetic spectra by calculating solar synthetic spectra, using the model 5800/4.50/0.0 ($T_{eff}$ / log g / [A/H]) of Gustafsson et al. (1975) and the "N" line list and comparing their results with observation. They also carried out these comparisons with three other line lists (referred to by them as O, P, and K), including the one used earlier for our studies of the 47 Tuc dwarfs. The wavelength regions 3840 - 3890 and 4250 - 4350 Å are shown in Figures 4a-c for the solar spectrum computed using the "N" line list, smoothed by a Gaussian with a FWHM of 0.05 Å, together with the observed solar flux spectrum (Kurucz et al. 1984). Logarithmic solar abundances of Fe, C, N, and O are assumed to be 7.52, 8.62, 8.00, and 8.86, respectively, on a scale of H = 12. We show these regions because of their relevance to the present investigation. On the basis of these comparisons, we believe our synthetic spectra are very satisfactory for studies of CN and CH in these 47 Tuc stars.

The resulting synthetic indices for three sets of C, N, and O abundances are given in Table 4. These new indices are compared, over the luminosity range of interest, to those from our previous work in Figure 5. Note that while we have shifted the model grid slightly, and greatly improved our input data, very little change in S(3839) is evident. The case with $s_{CH}$ is similar - the relative behavior of the band strength measurements are almost identical. The offset between the two sets of CH indices is due to improved atomic data and the inclusion of continuous CH-opacity, which essentially changes the zero-point defined by the flux in the continuum bandpass.

The model locus for C, N, and O in the solar ratios ([C,N,O/A] = 0) formed by a linear interpolation between grid points is shown in Figure 3, as is a comparison with non-solar ratios. A zero-point shift in the synthetic indices (0.14) has also been applied - the observed spectra have not been flux calibrated and therefore possess an additional instrumental response which results in such a shift. The synthetic spectra computed with [C/A] = –0.21 and [N/A] = +0.41 represent the maximum CN absorption that can be achieved via the CN-cycle of H-burning having started with C, N, and O in the solar ratio (see Smith & Bell 1986). The third set with [C/A] = –0.3, [N/A] = +0.76, and [O/A] = –0.2 depicts the effects of a moderate (but arbitrary) mix of C→N and O→N processed material to the stellar surface as in Smith, Bell & Hesser (1989). It is evident from Figure 3, that the presence of C→N cycled material alone is insufficient to account for the observed range in CN band strengths. Likewise, a significant fraction of the atmospheres must have been processed through the O→N cycle if one wishes to resort to an origin for the CN-strong stars which involves interior processing. Identical conclusions have been reached regarding brighter, more evolved stars in this cluster (see Smith et al. 1989). Additional calculations show that other differences, such as a range in $^{12}C/^{13}C$ ratios, are unlikely to appreciably affect CN band strengths among these stars (e.g., for the $T_{eff}$ = 5820, log g = 4.05 model, a change of $^{12}C/^{13}C$ from 89 to 10 results in an increase in S(3839) of 0.003).

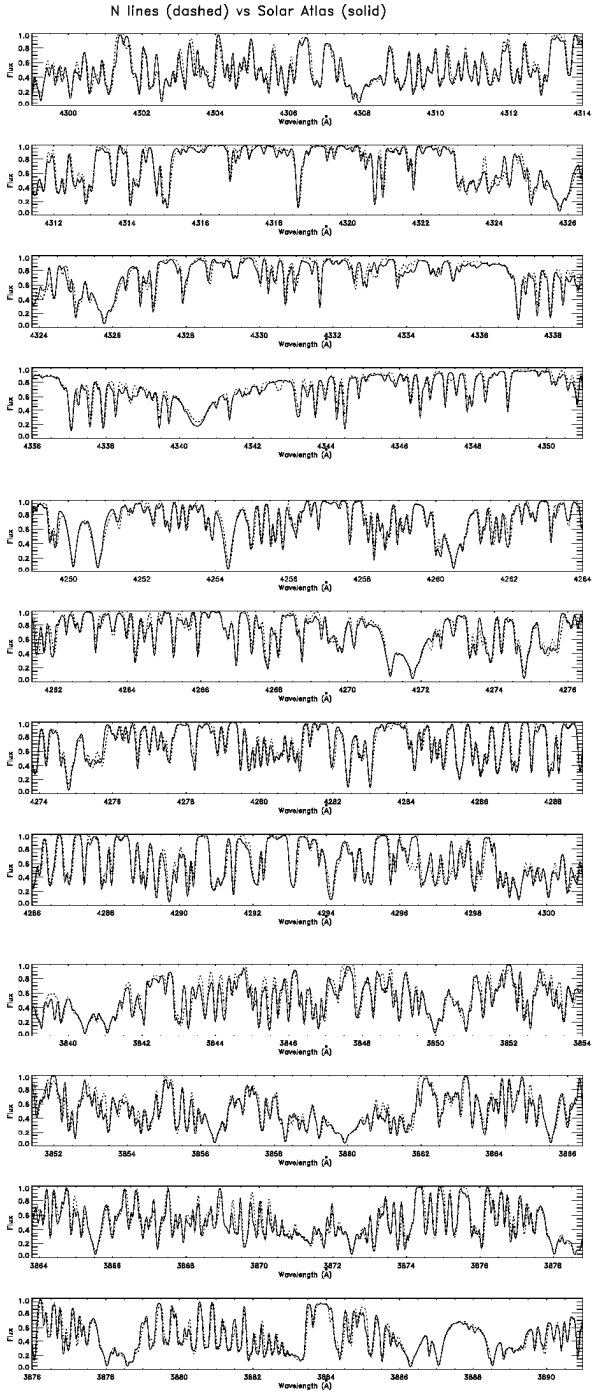

Figures 4a-c: Synthetic spectra generated from the solar model (5800/4.5/0.0) are compared with the solar flux atlas for the CN and CH regions of interest. The fit using the new molecular data is excellent.

An S(3839) residual, denoted as δS(3839), is defined as the difference between the observed S(3839) value of a star and that computed for a model atmosphere of the appropriate V magnitude and [C,N/A] = 0 (i.e., δS(3839) is a CN absorption excess from which temperature effects have been removed). A histogram of these residuals is shown in Figure 6 for the program stars fainter than V = 17.3, in which the bin size is equal to the average error in S(3839), 0.05.

The CN distribution among the red giant branch stars of 47 Tuc was found by Norris & Freeman (1979) to be bimodal, based on a much larger sample of stars than that of the present survey. The observed distribution of points in Figure 6 is also consistent with a bimodal distribution in the sense that is does exhibit two local maxima in the δS(3839) bins 0.075 - 0.125 and 0.325 - 0.375. However, the number of stars in each of these bins is only 5 and 4, respectively. A Kolmogorov-Smirnov test results in a 70% probability that the observed distribution was not drawn from a continuous sample (i.e. a uniform spread in CN band strengths). Consequently, perhaps the most that can be said about the shape of the CN distribution from the present data is that it is not inconsistent with a bimodality. The observations provide no obvious evidence of a difference between red giant branch and main-sequence turnoff CN distributions.

The $s_{CH}$ index, plotted versus V magnitude in Figure 7, reveals evidence for a range of CH band strengths among the turnoff stars (equal to approximately 0.07, or slightly more than 3σ). Briley et al. (1991) found a similar spread in G-band strengths among their program stars. As with Figure 3, the CN-strong stars have been marked with filled circles. It is apparent from this plot that the CN-strong stars tend to possess somewhat weaker CH bands than do their CN-weak counterparts. Indeed, with but one possible exception, all the CN-weak stars fall in the upper portion of the plot. Also plotted in Figure 7 are the $s_{CH}$ band strengths corresponding to the synthetic spectra represented in Figure 3. As with the CN indices, a zero-point shift has been applied (equal to –0.06) to the calculated indices. The range in CH band strengths is consistent with star-to-star [C/H] variations of at least a few tenths of a dex, as have also been observed among brighter stars (e.g., the lower giant branch stars in Smith et al. 1989).

A more instructive graph is shown in Figure 8, where the $s_{CH}$ index residuals ($\delta s_{CH}$) are plotted against δS(3839). $\delta s_{CH}$ is defined to be the difference between the observed $s_{CH}$ value of a star and that computed from a model atmosphere with C and N in the solar ratio with respect to iron. Thus Figure 8 represents a comparison of CN and CH band strengths from which the temperature component has been removed. As expected from Figure 7, the CN-weak stars appear to have somewhat stronger G-band strengths. This is also apparent in Table 5, where the statistics for the CN-strong and CN-weak stars are given. While the average error of a single measurement is comparable to the average separation in $\delta s_{CH}$ between the CN-strong and weak stars, the errors about the means of the two groups are much smaller, implying that the spread is real.

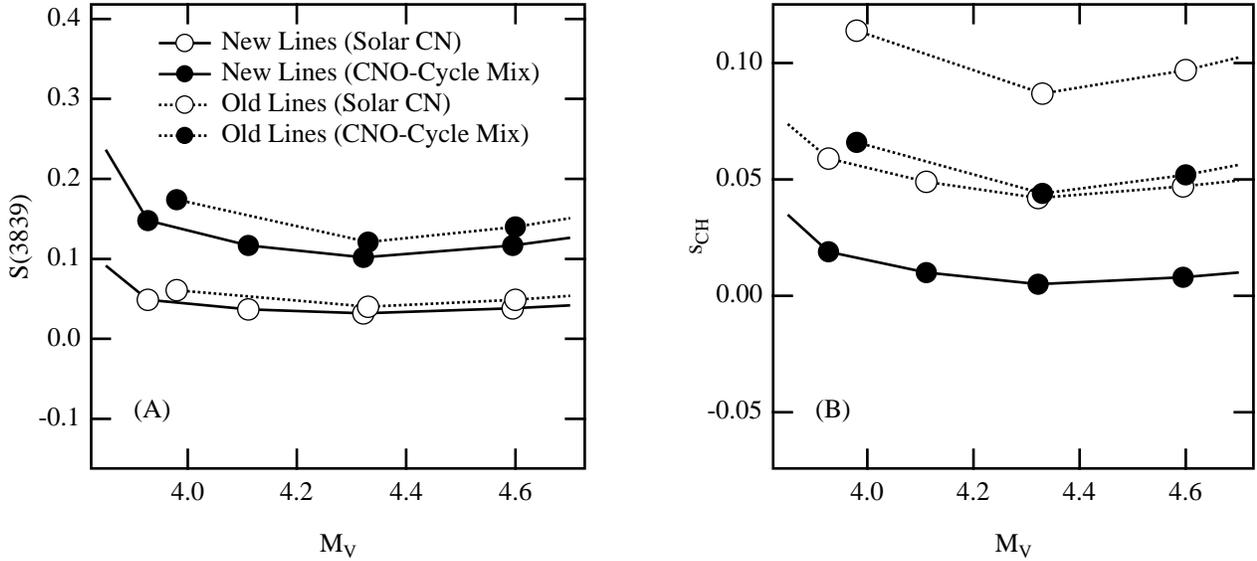

Figure 5: The indices computed using the new grid of models and line list are compared with those of Briley et al. (1991) over the luminosity range of interest. The relative sensitivities of the indices to changes in composition are similar.

Both Figures 7 and 8 and Table 5 suggest that an anticorrelation exists, on average, between the CN and CH band strengths of the 47 Tuc main-sequence turnoff stars. This substantiates and strengthens the evidence published by Bell et al. (1983) and Briley et al. (1991) regarding a CN-CH anticorrelation among pre-giant-branch stars in 47 Tuc, and similar findings in NGC 6752 by Suntzeff (1989) and Suntzeff & Smith (1991). A natural interpretation is that the carbon and nitrogen abundances of the turnoff stars tend to be anticorrelated, with observed CN enhancements being largely the consequence of enhanced nitrogen abundances.

Since the new data in a randomly selected sample of 47 Tuc turnoff stars yield ranges in CN/CH strengths similar to those found by Briley et al. (1991) in a sample designed to probe the extrema, we may prima facie adopt the latter's conclusions, namely: 1) the overall range in C and N abundances required to produce the main-sequence observations is similar to that required to explain data for luminous red giants; and 2) if one were to insist on a mixing origin for the dispersion, additional conversion of O→N would be required to maintain equality of C+N+O among all stars.

These 47 Tuc main-sequence turnoff star observations, and our earlier observations, support the integrated light models of Tripicco & Bell (1992). It appears that in such models it is necessary to include spectra of CN-strong stars at the turnoff, as well as throughout the giant branch, in order to match the strength of the CN bands in the integrated light of 47 Tuc.

## 4. DISCUSSION

The patterns of CN and CH differences observed among the main-sequence turnoff stars in 47 Tuc are similar to those found among the bright red giants (Norris & Freeman 1979), the red horizontal branch stars (Norris & Freeman 1982), and the subgiants (Smith et al. 1989 and Briley et al. 1989) in this cluster. Therefore, *at least some component of the C and N abundance inhomogeneities observed among the red giants in 47 Tuc appears to have been established prior to the commencement of evolution up the red giant branch.* Despite

TABLE 5:
Statistics of CN-strong and CN-weak Groups

| Group | n | $<\delta S(3839)>$ | $<\sigma$ (1 obs.)$>$ | $\sigma$ (mean) | $<\delta s_{CH}>$ | $<\sigma$ (1 obs.)$>$ | $\sigma$ (mean) |
|---|---|---|---|---|---|---|---|
| CN-strong | 12 | 0.301 | 0.060 | 0.017 | -0.096 | 0.027 | 0.008 |
| CN-weak | 8 | 0.122 | 0.037 | 0.013 | -0.061 | 0.031 | 0.011 |

ANTICORRELATED CN AND CH VARIATIONS ON THE 47 TUC MAIN-SEQUENCE TURNOFF 9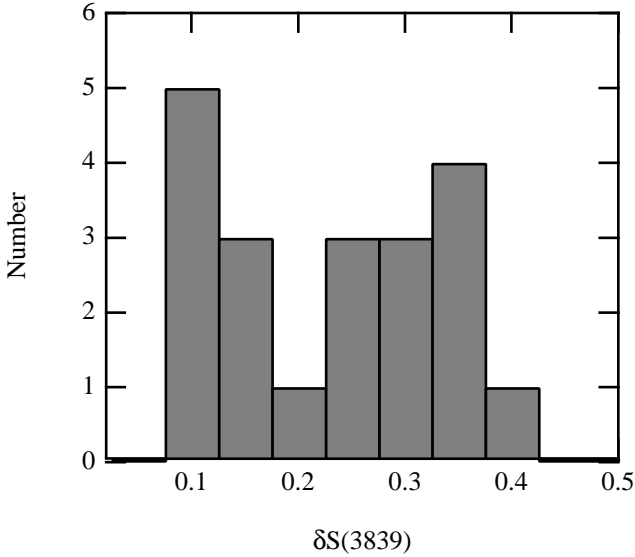

Figure 6: The histogram of CN-band strength excesses, δS(3839), as measured relative to models with C/N/O in the solar ratio. The bin sizes are 0.05, which is the average error in the S(3839) measurements.

the demonstration by Langer, Hoffman, & Sneden (1993) that a number of properties of giant-branch abundance inhomogeneities could feasibly be produced during normal giant-branch evolution, our observations indicate that there must be a more primitive cause for at least *some* of the star-to-star C and N differences among the stars in 47 Tuc.

One way in which early N enhancements could have been established among the main-sequence stars of 47 Tuc is via the accretion of N-rich material that had been ejected into the intracluster environment by intermediate mass stars. As pointed out by Hesser et al. (1985), the amount of accreted N would not need to be very high to produce a surface N enhancement for a main-sequence star; because these stars have relatively shallow convection zones, the accreted N would be distributed throughout only a small percentage of the stellar interior. However, when such a superficially enriched main-sequence star evolves into a red giant, the convective envelope deepens, and the surface N excess would then be distributed throughout a much larger percentage of the interior mass, resulting in a decrease of the surface N abundance. Consequently, if the N enhancements of the 47 Tuc main-sequence stars were to be produced by the accretion of N onto their surfaces, then it would be expected that the red giants of 47 Tuc should have much smaller N abundances than the N-rich main-sequence stars. This is not observed to be the case, as both the ratio of CN-strong to CN-weak stars and the range of CN band strengths appears consistent with those found among the evolved giants. Therefore, such selective accretion seems an unlikely explanations for the CN variations in 47 Tuc.

Another alternative might be to suggest that the N-enriched main-sequence stars obtained their excess N via mass transfer from other stars in the cluster during interactions. This scenario however, suffers from several shortcomings. One of these is similar to that noted above; any mass transfer of N-rich material would enrich only a relatively shallow outer convective zone of a main-sequence star, leading to the problem noted in the previous paragraph. In addition, our spectroscopically observed sample of main-sequence stars lies at radii between about $4r_c$ and $8r_c$, and it is uncertain as to whether such stars would spend sufficient time near the dense cluster core where stellar interactions would be most likely. We note however the observations of Norris & Freeman (1979) which demonstrate a significant difference in the distribution of CN band strengths between 47 Tuc giants with r < 3 and r > 6 arc min, with the stars radially inward tending to be CN strong. A similar result has also been obtained by Paltoglou (1991) and it would be of interest to know whether this behavior is also seen among the main-sequence stars.

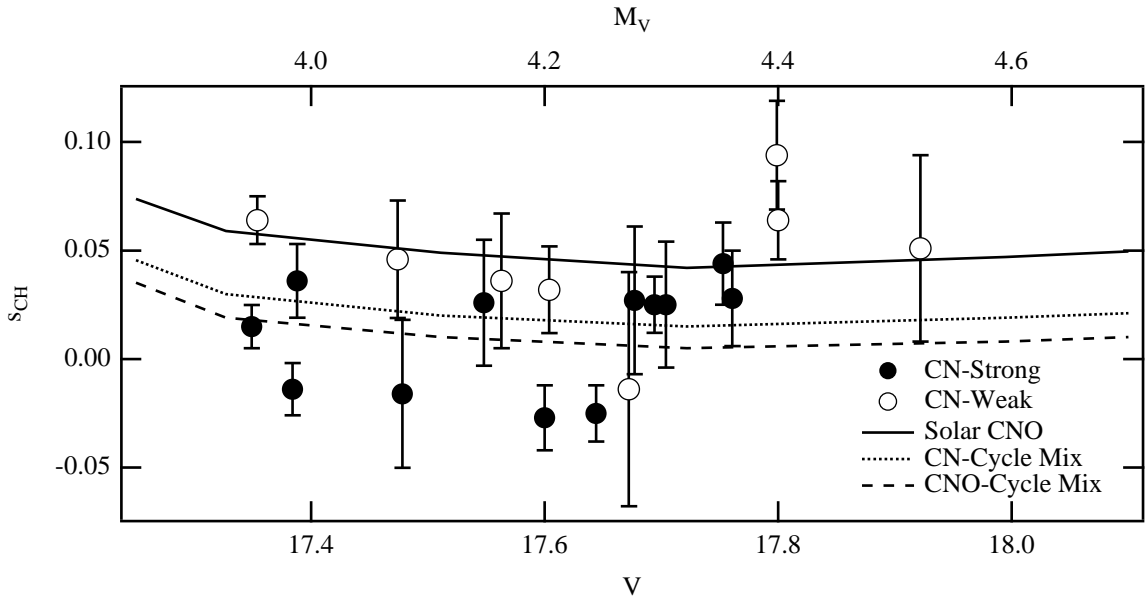

Figure 7: CH (G-band) strength indices are plotted for the stars in Figure 3. Note that the CN-weak stars (open markers) tend to have stronger G-bands than do their CN-strong counterparts (solid markers), implying a C vs N abundance anticorrelation. Also plotted are the corresponding synthetic indices from the models of Figure 3.

As discussed by VandenBerg & Smith (1988), the observation of CN differences as we have found among main-sequence turnoff stars is very difficult to reconcile with any hypothesis which seeks to attribute the production of surface nitrogen abundance enhancements among the CN-strong stars to the dredge-up, by whatever process, of C→N processed material from the interior of the star. The interior models of such turnoff stars presented by VandenBerg & Smith (1988) show that the region of C→N burning is contained within the central H-burning region, and so is located in a region through which there is a significant molecular weight gradient. Such gradients are thought to hinder the action of meridional circulation, one of the main theoretical contenders for mixing in cluster red giants (Sweigart & Mengel 1979). In addition, the proximity of the H-burning and C→N processing regions provides another problem for mixing, since any process capable of bringing newly synthesized nitrogen up from the CN-processing region of the star would also bring down fresh hydrogen into the H-burning region, thereby prolonging the main-sequence lifetime. Such mixing, if it occurred within some stars and not others, as the observations imply, would also be expected to produce a range in mass and color among the turnoff stars, as indicated by the evolutionary tracks of VandenBerg & Smith (1988). As noted by them, however, the main-sequence turnoff in the CMD of Hesser et al. (1987) is no wider than is explicable in terms of (the small) observational errors.

For such reasons, ab initio abundance differences between 47 Tuc stars seem a less forced explanation for the CN and CH differences among main-sequence turnoff stars, which suggests that 47 Tuc may have been initially chemically inhomogeneous in C and N. Such inhomogeneities might have resulted from the nucleosynthetic activity of stars more massive than those still currently evolving within the cluster (Cottrell & Da Costa 1981, Smith & Wirth 1991, and references therein).

Alternatively, to account for the narrow evolutionary sequences of the 47 Tuc CMD, as well as the C and N differences among the turnoff stars, by some form of stellar-interior process that has acted within the cluster stars at some point prior to the turnoff, it seems necessary to hypothesize the following interior phenomena: a) Some transport mechanism, differing in efficiency among the 47 Tuc main-sequence stars, is required to bring nucleosynthesized material from their interior to the surface. b) A region of C→N conversion must exist exterior to the H→He burning core within at least some of the main-sequence stars in 47 Tuc (so as to allow access by the hypothesized transport mechanism to regions of N enhancement, but not to the H-depleted regions where the main source of energy production by the p-p chain of hydrogen fusion is occurring). Neither of these phenomena are included in published interior models of globular cluster main-sequence stars, nor are they a natural consequence of such models.

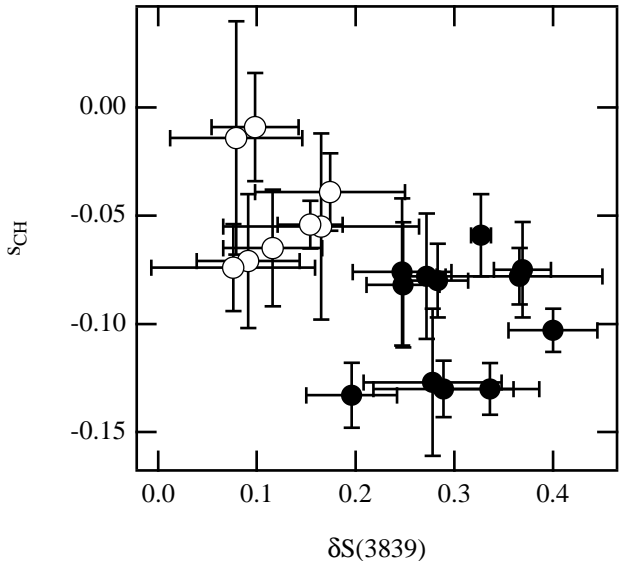

Figure 8: CN and CH band strength excesses are plotted against each other for the stars of Figure 6.

A cautionary note remains: previous surveys of similar dwarf stars in the field, such as those of Clegg, Tomkin & Lambert (1981); Tomkin & Lambert (1984); Laird (1985); and Carbon et al. (1987), have revealed essentially no significant scatter in [N/Fe] at any given [Fe/H] from –2 to +0.3. Although there does appear to be a small population of N-rich stars (5%) in the samples of Laird (1985) and Carbon et al. (1987), the Li study of Spite & Spite (1986) implies that the N-enhancements could not have been the result of deep dredge-up in at least four of the five stars examined. Moreover, recent observations of two of these stars by Beveridge & Sneden (1994) demonstrate enhanced s-process abundances, further implying that these stars are not analogous to the N-rich cluster dwarfs. Thus we are confronted with the additional constraint that if a mixing process is postulated to explain the present observations, it must be rather unique to the interiors or structures of the globular cluster stars.

## 5. CONCLUSIONS

The present results confirm the existence of star-to-star CN and CH variations among an unbiased sample of unevolved stars in 47 Tuc. The distribution of CN band strengths is consistent with a bimodal distribution and the ratio of CN-strong to CN-weak stars is similar to that found among the subgiant, red-giant, horizontal-branch, and asymptotic-giant branch stars. A CN vs CH anticorrelation, similar to those observed among the more evolved cluster stars, has also been verified. These results imply that the mechanism largely responsible for these variations was in operation at a very early point in the evolution of the 47 Tuc stars, before the onset of H-shell burning and giant branch ascent.

The range of CN and CH band strengths are in agreement with those observed among similar stars by Briley et al. (1991), and their conclusions requiring either the dredge-up of ON-cycled material to the stellar surface, or a non-solar initial C/N ratio with deep mixing of the CN-weak stars in order to maintain constant C + N + O, still hold. Given that no current model of stellar structure hypothesizes such a mechanism, the simplest scenario would seem to be a primordial source (with the added caution that although signs of CN or ON-processed material may be present in the atmosphere of a cluster star, this does not necessarily imply that the star was actually the site of the nucleosynthesis).

We would especially like to thank H. Tirado and the staff of the Cerro Tololo Inter-American Observatory for the unique instrument which made this project possible and Dr. Allen Sweigart for his comments on the manuscript. We would also like to thank Dr. Nick Suntzeff for the careful attention he paid to this manuscript during the review process and for his helpful suggestions. Part of this work was sponsored by a Texas Advanced Research Program grant to MMB. RAB wishes to acknowledge support under NSF grants AST 89-18461 and AST 91-22361.